\documentclass[prd,
 reprint,
 amsmath,amssymb,
 aps,nofootinbib
]{revtex4-2}

\usepackage{bm}

\PassOptionsToPackage{linktocpage}{hyperref}
\usepackage[hyperindex,breaklinks,hidelinks]{hyperref}
\usepackage{enumitem}
\usepackage{slashed}
\usepackage[dvipsnames]{xcolor}

\usepackage{subcaption}

\usepackage{tikz}
\tikzset{every picture/.style={line width=0.75pt}} 
\usetikzlibrary{shapes.geometric}
\usetikzlibrary{decorations.markings}
\usepackage{orcidlink}
\usepackage{comment}
\usepackage{array}
\usepackage{mathtools}

\usepackage{etoolbox}

\newcommand{\Lag}{\mathcal{L}}

\newcommand{\ket}[1]{\ensuremath{\left|\, #1\right>}}

\def\d{\text{d}}

\begin{document} 

\title{The Strong-$CP$ Problem and its Gauge Axion solution \\ as Evidence for Fundamental Strings}


\author{Gia Dvali} 
\email{georgi.dvali@lmu.de}

\author{Lucy Komisel\,\orcidlink{0009-0008-1578-588X}} 
\email{lucy.komisel@mpp.mpg.de}

\author{Otari Sakhelashvili} 
\email{o.sakhelashvili@lmu.de}

\author{Anja Wachowitz\,\orcidlink{0009-0005-0920-379X}} 
\email{anja.stuhlfauth@mpp.mpg.de}

\affiliation{Arnold Sommerfeld Center, Ludwig-Maximilians-Universit\"at, Theresienstra{\ss}e 37, 80333 M\"unchen, Germany}
\affiliation{Max-Planck-Institut f\"ur Physik, Boltzmannstr. 8, 85748 Garching, Germany}

\date{\today}

\begin{abstract} 

The topological susceptibility of the QCD vacuum provides an understanding of $\theta$-vacua as vacua of a Chern-Simons gauge theory. 
In this way, it gives an immediate proof of the physicality of the boundary $\theta$-term.  
This makes the essence of the strong-$CP$ puzzle very transparent and offers a solution in form of the gauge axion, which has exact quality.    
This axion represents an intrinsic part of the QCD gauge redundancy without any reference to an anomalous global symmetry. 
It is a two-form transforming under the QCD gauge symmetry.  
Due to its pure gauge nature, the gauge axion represents a powerful tool to monitor physics of $\theta$-vacua in various regimes. 
Unlike the ordinary Peccei-Quinn axion, which is UV-completed into a Goldstone phase of a complex scalar and thereby suffers from the quality problem, the gauge axion is UV-completed directly into a fundamental theory of gravity.   
We study the domain wall and string structure of the gauge axion and show that the strings sourcing it must be a part of this fundamental theory.  
We thus observe that the absence of the axion quality problem motivates the presence of fundamental strings. 
This provides a new argument for a connection between the axion and gravity.  
\end{abstract}

\maketitle

\section{Introduction} 
It is well-known that Yang-Mills gauge theories possess a continuum of vacuum states, labelled by a $CP$-violating parameter $\theta$ \cite{Callan:1976je,Jackiw:1976pf}. 
In the context of QCD, this parameter is physically measurable via the electric dipole moment of the neutron (nEDM) \cite{Baluni:1978rf,Crewther:1979pi}, resulting in the experimental upper bound $\theta  \lesssim 10^{-9}$ \cite{Baker:2006ts, Pendlebury:2015lrz, Graner:2016ses}\footnote{For compactness of the present discussion, we assume that the contribution from the phase of the determinant of the quark mass matrix has been taken into account in $\theta$.}.
The smallness of $\theta$ is the source of the so-called strong-$CP$ puzzle, formulated as a naturalness problem. 
  
The most celebrated solution to this problem is the Peccei-Quinn (PQ) solution \cite{Peccei:1977hh,*Peccei:1977ur}, based on removing $\theta$ by a global, anomalous, chiral symmetry $U(1)_{PQ}$. 
This symmetry is spontaneously broken, resulting in a pseudo-Goldstone particle, the axion \cite{Weinberg:1977ma, *Wilczek:1977pj}. 
However, the PQ formulation is sensitive to arbitrary continuous, deformations of the theory by operators, which explicitly break the $U(1)_{PQ}$-symmetry. 
This is commonly referred to as the \textit{axion quality problem}.  
  
For some time, it has been understood that the topological susceptibility of the vacuum (TSV) provides a powerful tool to understand the physics of $\theta$-vacua in a pure-gauge language of the Chern-Simons (CS) $3$-form \cite{Dvali:2005an, Dvali:2005zk, Dvali:2005ws,  Dvali:2007iv, Dvali:2013cpa, Sakhelashvili:2021eid, Dvali:2022fdv,  Dvali:2023llt}.  
Notice that this formulation is universally applicable to arbitrary sectors with a non-trivial TSV, including the electroweak one \cite{Dvali:2024zpc, Dvali:2025pcx} as well as 
gravity \cite{Dvali:2005an, Dvali:2022fdv, Dvali:2024dlb}.
 
It was also realized that this picture offers a pure-gauge formulation of the axion solution in terms of a $2$-form field $B_{\mu\nu}$ \cite{Dvali:2005an, Dvali:2005zk, Dvali:2013cpa, Sakhelashvili:2021eid, Dvali:2022fdv,  Dvali:2023llt}. 
The advantage of this formulation is that the axion emerges as an internal part of the QCD gauge redundancy, without the need for any anomalous global symmetry. 
Due to this, the gauge axion solution has exact quality. 
That is, the  solution is insensitive to arbitrary continuous deformations of the theory to all orders in the operator expansion. 
     
While at the level of the low-energy effective theory (EFT) the gauge axion solution is dual to a pseudo-Goldstone axion of exact quality, this duality is broken by the UV-completion. 
The PQ axion is UV-completed into a phase of a complex scalar, which is the source of its quality problem. 
In contrast, the gauge axion, which possesses exact quality, has no known completion within a renormalizable EFT and, instead, has to be UV-completed directly within the UV-theory of gravity.         
        
In the present paper, we shall deepen the connection between the strong-$CP$ puzzle and the gauge redundancy using the gauge axion as a monitoring tool. 
we will first show that the TSV offers a direct proof of the physicality of the $\theta$-parameter. 
Then we show that the gauge axion makes it possible to explicitly scan various regimes of the theory. 
Namely, we demonstrate how the $\theta$-vacua continuously re-emerge in a limit in which the gauge axion decouples.  
  
Next, we study the structure of the gauge axion string-wall systems. 
We observe that, while domain walls are fully accountable as explicit solutions in the low-energy EFT, axionic strings must be fundamental. 
That is, via the gauge axion, the solution of the quality problem of the low-energy theory is directly linked to the existence of fundamental string entities sourcing the gauge axion. 
  
This finding is remarkable from at least two perspectives. 
First, it resonates with previous conclusions, reached from very different considerations that gravity requires an exact quality axion, for each gauge sector of the theory, by the consistency of its formulation \cite{Dvali:2022fdv,  Dvali:2023llt, Dvali:2024dlb}.  
   
Second, it is known that fundamental superstrings, from a low-energy perspective, can play the role of cosmic strings sourcing axion-like two-forms $B_{\mu\nu}$ \cite{Witten:1985fp}.
More discussions on cosmic $F$- and $D$- strings were given in \cite{Dvali:2003zj, Copeland:2003bj}.      
   
Whether any of these particular fields can play the role of the QCD axion, is a different matter.  However, the key lesson we take from the above is that the known setups, in which an axion is sourced by fundamental strings, must incorporate gravity. 
In this light, it is indicative that the same conclusion is reached in our bottom-up analysis, which is based solely on the gauge redundancy and the requirement of an exact quality axion. 

\section{Topological Formulation of the Strong CP Problem}

It is well-known that, in the absence of massless quarks, QCD  contains a continuum of vacuum states $\ket{\theta}$, which obey a superselection rule \cite{Callan:1976je, Jackiw:1976pf}.   
The choice of vacuum is accounted for by the boundary term, 
\begin{align} \label{Stheta} 
    S_{\theta}\, = \frac{\theta}{16\pi^2}\int \d^4x\, \,  G_{\mu\nu}\tilde{G}^{\mu\nu},
\end{align}
with $G_{\mu\nu}\tilde{G}^{\mu\nu} \equiv  \epsilon^{\mu\nu\alpha\beta} {\rm tr} G_{\mu\nu}G_{\alpha\beta}$, where $G_{\mu\nu}$ is the $SU(3)_C$ field-strength of the gluon field matrix $A_{\mu} \equiv A_{\mu}^aT^a$, with 
$T^a, ~a=1,2,...,8$ the generators. 
 
This can be represented by a total derivative,  
\begin{equation} \label{dCS}
  G \tilde{G}  = 
   \epsilon^{\mu\nu\alpha\beta} \partial_{\mu} C_{\nu\alpha\beta},
\end{equation}
where, 
\begin{equation} \label{CS}
   C_{\mu\nu\alpha}  \equiv {\rm tr} \left(A_{[\mu}\partial_{\nu}A_{\alpha]} + \frac{2}{3}A_{[\mu}A_{\nu}A_{\alpha]}\right) \,,
\end{equation}
is the CS $3$-form.
Under the QCD gauge transformation, $A_{\mu} \rightarrow U A_\mu U^{\dagger} + U\partial_{\mu}{U^{\dagger}}$, with $U = exp(iT^C\omega^C)$, the 
$3$-form shifts as 
\begin{align} \label{ShiftQCD}
    C_{\mu\nu\alpha} \rightarrow C_{\mu\nu\alpha} + \partial_{[\mu}\Omega_{\nu\alpha]}\,,
\end{align}  
 with $\Omega_{\mu\nu} \equiv A_{[\mu}^a\partial_{\nu]}\omega^a$. 

The physicality of the $\theta$-term is directly linked to the TSV, 
\begin{align}
\label{EEcorr}
    FT \langle G\tilde{G},  G\tilde{G}\rangle_{p\to 0} \equiv \lim_{p \to 0}\int \d^4 x\, e^{ipx} \langle T [G\tilde{G}(x), G\tilde{G}(0)]\rangle,
\end{align}
where $T$ stands for time-ordering and $FT$ stands for the Fourier transformation.
In pure glue, the above correlator is a non-zero constant, set by the QCD scale $\Lambda^4$.  
For the time being, we shall work in units of this scale.

The $\theta$-term is physical, if the above correlator is non-zero and vice versa \cite{Witten:1979vv, Veneziano:1979ec}.  
Although, by now this connection is common knowledge, we will provide a simple derivation for completeness.  
For this, as a first step, we rewrite \eqref{Stheta} as a boundary term, 
\begin{align} \label{ST} 
   \theta \int \d^4x \,G_{\mu\nu}\tilde{G}^{\mu\nu}(x) =  \int \d^4x\  C_{\mu\nu\alpha}(x) {\mathcal J}^{\mu\nu\alpha} (x) ,
\end{align}
where 
\begin{eqnarray} \label{JBoundary} 
  && {\mathcal J}^{\mu\nu\alpha}(x) \, \equiv \,
 \nonumber \\ 
  && \theta  \, \int_{2+1} 
  \d^3\xi \, \delta(x - X(\xi)) \, \frac{\partial X^{\mu} \partial X^{\nu} \partial X^{\alpha}}
  {\partial \xi^a \partial \xi^b \partial \xi^c} \epsilon^{abc} \,, 
\end{eqnarray}
is a current which describes a $(2+1)$-dimensional boundary with world-volume coordinates $\xi^{a},~a,b,c\, = \,0,1,2$, and the embedding spacetime coordinates $X^{\mu}(\xi^a)$.    
The quantity $\epsilon^{abc}$ denotes the totally antisymmetric tensor of the world-volume. 
The boundary current is trivially conserved, $\partial_{\mu}{\mathcal J}^{\mu\nu\alpha}(x)=0$. 
  
Next, let us consider an arbitrary effective operator ${\mathcal O}(x)$ describing a physical process that interacts with the CS current $C_{\mu\nu\alpha}(x)$. 
In general, ${\mathcal O}(x)$ is a composite operator, made up of the fields participating in the process.   
We shall put no restriction on this composition. 
An example of such a process can be a measurement of the nEDM, but any other physical process would be equally valid. 
Obviously, gauge invariance and the topological nature of the CS current imply that the source must be conserved. 
 
Correspondingly, the coupling can have the form $C^{\mu\nu\lambda} \epsilon_{\mu\nu\lambda\sigma}\partial^{\sigma} {\mathcal O}(x)$, or equivalently, 
\begin{eqnarray} \label{OGG} 
    \int \d^4x\, \, {\mathcal O}(x)\, G_{\mu\nu}\tilde{G}^{\mu\nu}(x). 
\end{eqnarray}
The above operator is of course not a boundary term and describes a physical process. 
The question is, whether the process will depend on $\theta$. 
 
Naively, it will not, since $\theta$ is a boundary term, whereas the above operator describes local bulk physics. 
However, this is not the case. 
To see this, we evaluate the $S$-matrix element that connects eq. \eqref{ST} with the boundary $\theta$-term, 
\begin{eqnarray}
\label{S-matrix}
    S = \int \d^4x \d^4x' \,
    {\mathcal O}(x)\langle  G_{\mu\nu}\tilde{G}^{\mu\nu}(x)
    C_{\alpha\beta \gamma} (x')\rangle {\mathcal J}^{\alpha\beta\gamma}(x') \,, 
\end{eqnarray} 
where the brackets $\langle ...\rangle$ stand for the contraction.  
Taking into account eqs. \eqref{EEcorr},\eqref{ST} and \eqref{JBoundary} and integrating over $x'$, the above expression can be reduced to 
\begin{equation}
\label{Smatrix}
   S =   \frac{\theta}{16\pi^2}\int \d^4x \,{\mathcal O}(x) \,. 
\end{equation} 
In other words, despite the boundary nature of the 
$\theta$-term \eqref{ST}, the matrix element of the bulk process, described 
by the operator ${\mathcal O}(x)$, is explicitly $\theta$-dependent.
This proves the physical nature of the $\theta$-vacua. Notice that, the (\ref{Smatrix}) also proves that $\langle G_{\mu\nu}\tilde{G}^{\mu\nu} \rangle \propto \theta$.

The connection  between \eqref{EEcorr} and the physical nature of $\theta$ can be understood in the language of a $3$-form gauge theory \cite{Dvali:2005an}. 
In order to formulate this description, first notice that eq. \eqref{EEcorr} implies a massless pole in the correlator of the 3-form, $\langle C, C \rangle$. 
The existence of this pole was noticed by L\"uscher \cite{Luscher:1978rn}, without connecting it to the strong-$CP$ problem. 

More precisely, from \eqref{EEcorr} and the existence of a mass gap in the propagating degrees of freedom of QCD, it follows directly that the  K\"all\'en-Lehmann spectral representation of the 3-form correlator includes a physical pole at $p^2=0$
\cite{Dvali:2005an, Dvali:2017mpy, Dvali:2022fdv, Dvali:2023llt},    
\begin{equation}
\label{CCcorr}
    \langle C ,C \rangle =  \frac{1}{p^2} + \sum_{m\neq 0} 
    \frac{\rho(m^2)}{p^2 - m^2} \,, 
\end{equation} 
where $\rho(m^2)$ is a normalized spectral function.

Notice that, as is the case for any other massless gauge field, the massless pole in the correlator is gauge-invariant. 
This can be seen in several different ways, in particular, by computing a gauge invariant force mediated between the two conserved sources~\cite{Dvali:2024zpc}.
  
In pure glue, the massive poles are separated by a mass gap of order the QCD scale.  
Correspondingly, the vacuum structure of the theory is \textit{completely} determined by the massless pole ~\cite{Dvali:2005an}. 
This entry corresponds to a massless $3$-form field, which we denote by the same symbol $C_{\mu\nu\alpha}$.     
The corresponding gauge invariant field strength shall be denoted by $E = \epsilon^{\mu\nu\alpha\beta} \partial_{\mu}C_{\nu\alpha\beta}$.   
The vacuum structure of the theory is then fully determined by the effective Lagrangian
\begin{equation} \label{Kfunction1}
    L \, = \, {\mathcal K}\left(E\right) \,, 
\end{equation} 
where $\mathcal{K}(E)$ is an analytic function of $E$ \cite{Dvali:2005an}, which does not depend on derivatives of $E$. 
 
All higher derivative contributions to the low energy theory, including high-derivative contributions from the massive poles in \eqref{CCcorr}, are irrelevant for the vacuum structure as they vanish in the zero momentum limit. 
This is an exact statement. 
The vacuum is thus determined by the effective equation
\begin{equation} \label{Kfunction2}
    \partial_{\mu} \left [ \frac{\partial {\mathcal K}(E)}{\partial E} \right ] = 0\,, 
\end{equation} 
which is solved by an arbitrary constant $E = \mathrm{const.}$. 
Since there exist no massless charges sourcing $C$, this value of $E$ cannot be changed. 
Thus, the vacua obey a superselection rule. 
Since $\langle  E \rangle  = \langle G\tilde{G} \rangle $, the connection to the $\theta$-vacua is obvious. 
   
From the above, it is clear that, in order to eliminate the $\theta$-vacua, it is necessary to eliminate the massless pole from \eqref{CCcorr}. 
By gauge invariance, this is only possible if the $3$-form becomes massive.
Thus the solution of the strong-$CP$ puzzle requires a Higgsing of $C$ 
\footnote{The above indicates that the smallness of $\theta$ cannot be justified by a symmetry, since the choice of the vacuum (or any other solution of the theory) cannot be restricted by the symmetry that it breaks. 
This point has been made in a number of papers \cite{Dvali:2005zk, Dvali:2007iv, Dvali:2022fdv,  Dvali:2023llt} and was recently rediscussed in \cite{Kaplan:2025bgy, Benabou:2025viy}. 

However, in the present context we shall put the issue of restricting the parameter $\theta$ by a symmetry aside. We shall also not discuss 
the solutions to strong-$CP$ based on alternative to axion dynamical 
relaxation mechanisms, such as \cite{Dvali:2005zk, Kaloper:2025wgn}. 
Instead, we shall mainly focus on the physicality of the $\theta$-vacua and their elimination by the gauge axion. 
This is also justified by the point made in \cite{Dvali:2018dce, Dvali:2022fdv,  Dvali:2023llt, Dvali:2024dlb} that gravity demands an exact quality axion as a consistency requirement, regardless of naturalness considerations.}.
  
\section{Gauge Axion}

A massive pole in the $3$-form correlator corresponds to a massive pseudo-scalar.
Thus, the elimination of the $\theta$-vacua is directly linked to the existence of a massive pseudo-scalar particle. 
W.l.o.g. we can call this an "axion". 
Its origin remains a question.

In the standard Peccei-Quinn formulation \cite{Peccei:1977hh, *Peccei:1977ur}, the axion emerges as a pseudo-Goldstone degree of freedom \cite{Weinberg:1977ma, *Wilczek:1977pj} from the phase of a complex scalar field, which spontaneously breaks the anomalous global symmetry $U(1)_{PQ}$. 
However, this formulations suffers from the quality problem, as it is sensitive to continuous deformations of the theory by operators which explicitly break the $U(1)_{PQ}$-symmetry. 
Any such deformation restores the massless pole in the Chern-Simons correlator \eqref{CCcorr} \cite{Dvali:2005an, Dvali:2017mpy, Dvali:2022fdv, Dvali:2023llt}.

In the present paper we shall focus on a pure-gauge formulation of the axion \cite{Dvali:2005an}. 
In this picture, the action is introduced as an intrinsic part of the QCD gauge redundancy, without reference to any global symmetry. 
This formulation has two advantages.
First it demonstrates the physicality of the $\theta$-vacua directly from the gauge symmetry of QCD. 
Second, it provides an exact quality solution of the strong-$CP$ problem.

All we need is to introduce a 2-form $B_{\mu \nu}$, which shifts under QCD gauge transformations \eqref{ShiftQCD} as 
\begin{equation}
\label{shift}
    B_{\mu\nu} \, \rightarrow \, B_{\mu\nu} \, + \, \frac{1}{f}\Omega_{\mu\nu}\,,
\end{equation}
where $f$ is a scale. 
We shall refer to $B_{\mu \nu}$ as the gauge axion. 
This fully fixes the vacuum structure as well as the spectrum of the theory. 
The point is that $B_{\mu \nu}$ enters the action only via the following gauge invariant combination with the CS $3$-form of QCD: 
\begin{align}\label{eq:Stuckelberg}
    \tilde{C}_{\mu\nu\alpha} \equiv  C_{\mu\nu\alpha} - f\partial_{[\mu}B_{\nu\alpha]}\,.
\end{align}
The lowest order invariant is 
\begin{align} \label{MtermC}
    \frac{1}{f^2}\tilde C_{\mu\nu\lambda}\tilde C^{\mu\nu\lambda}\,,
\end{align}
and one can add arbitrary higher order invariants without affecting the vacuum structure. 
It is clear that $B_{\mu \nu}$ represents a Stückelberg field for the Chern-Simons $3$-form. 

Let us now see, that the above construction proves the physicality of the massless pole in pure QCD as well as its elimination in the presence of $B_{\mu \nu}$.   
In other words, both, the problem and the solution become transparent in one shot. 

Let us show that the gauge axion represents a gauge-invariant probe of the $\theta$-vacua. 
Namely, we show that the $\theta$-vacua are eliminated for finite $f$ and are recreated in the decoupling limit of the gauge axion $f \rightarrow \infty$. 
In order to see this, let us consider the vacuum structure determined by the effective Lagrangian 
\begin{equation} \label{Kfunction3}
   L = {\mathcal K}(E) \, +  \frac{1}{f^2}\tilde C_{\mu\nu\lambda}\tilde C^{\mu\nu\lambda}\,.  
\end{equation} 
For simplicity, we first take a bilinear kinetic function ${\mathcal K}(E)  \propto  E^2$.  
  
The resulting equation of motion is 
\begin{equation} \label{VacEQ1}
    \partial^{\mu} \partial_{[\mu}\tilde{C}_{\nu\alpha\beta]} \, 
    + \, m^2 \tilde{C}_{\nu\alpha\beta} \, = \, 0 \,.  
\end{equation} 
Taking the divergence, we get the following Proca-type constraint, 
\begin{equation} \label{Proca1}
  \partial^{\nu} \tilde{C}_{\nu\alpha\beta} \, = \, 0 \,.  
\end{equation} 
With this constraint, the equation of motion becomes, 
\begin{equation} \label{VacEQ2}
  \Box \tilde{C}_{\nu\alpha\beta} \, 
  + \, m^2 \tilde{C}_{\nu\alpha\beta} \, = \, 0 \,,  
\end{equation} 
 
For  $m \neq 0$, an unique vacuum solution corresponds to 
the zero field strength, 
\begin{equation} 
    \partial_{[\mu}\tilde{C}_{\nu\alpha\beta]}  \, = \, 0 \,.  
\end{equation} 
Or equivalently, $E = G\tilde{G} = 0$.  
This unique vacuum is nothing but the $CP$-conserving vacuum corresponding to $\theta \, = \, 0$. 
We thus see that for any finite value of $f$ the gauge axion eliminates all $CP$-violating vacuum states. 
The reason is that the Chern-Simons $3$-form is in the Higgs phase \cite{Dvali:2005an}. 
 
It is clear that the addition of arbitrary higher order operators cannot change this, since they cannot un-Higgs the $3$-form as long as $f$ is finite.  
This may create a false impression that the uniqueness of the vacuum persists in the limit $f \rightarrow \infty$. 
However, this is not the case, since in this limit $B_{\mu\nu}$ decouples and the $3$-form is un-Higgsed. 
Therefore in the decoupling limit the $\theta$-vacua reemerge. 
However, the limit is perfectly continuous. 
The continuity is in the relaxation time of states that become true $\theta$-vacua  for infinite $f$. 
That is, in order not to overlook the vacuum states, one must track all the states that become eternal in the limit $f\rightarrow \infty$~\cite{Dvali:2022fdv}.  
  
Let us track these vacua explicitly. 
For this, notice that, at finite $f$, there exists a continuum of  spatially uniform oscillatory solutions
$C_{\mu\nu\alpha} = \epsilon_{0 \mu\nu\alpha} C(t)$ 
where  
\begin{equation} \label{Cosc1}
    C(t)  = {\mathcal A} \sin(mt + \varphi),  
\end{equation} 
with arbitrary amplitude ${\mathcal A}$ and phase $\varphi$.
The expectation value of $G\tilde{G}$ oscillates accordingly as 
\begin{equation} \label{Cosc2}
    G\tilde{G} =  \d_tC(t) =  {\mathcal A} m \cos(mt + \varphi).  
\end{equation}  
Now, the recreation of the $\theta$-vacua takes place in the double scaling limit $f\rightarrow \infty$ (or, equivalently $m \rightarrow 0$)  with the product ${\mathcal A} m$ kept finite but arbitrary.
This product parameterizes the vacua which correspond to different fixed values of $G \tilde{G}$. 
Since in this limit the gauge axion decouples, the vacua obey a strict superselection rule.  
     
Notice that, for a larger value of the amplitude ${\mathcal A}$, the harmonic oscillator solution (\ref{Cosc1}) must be replaced with the solution $C(t)$ of the equation 
\begin{equation} 
    \d_t({\mathcal K}(\d_tC(t))') 
    + m^2 C(t) = 0,  
\end{equation} 
 where the prime denotes the derivative of the ${\mathcal K}$-function with respect to its argument. 
 The explicit form of the solution is unimportant.
The crucial thing is that the oscillation frequency $m$ goes to zero in the limit $f \rightarrow \infty$. 
In this limit, all oscillatory solutions freeze and become $\theta$-vacua marking the different strict superselection sectors of the theory. 

\section{The Gauge Axion as a Window to Fundamental Physics}

It can be shown that the gauge axion can be dualized into a pseudoscalar $a$ with the potential $V(a)$ fully determined by the kinetic function ${\mathcal K}$
\cite{Dvali:2005an, Dvali:2013cpa, Dvali:2017mpy},\cite{ Sakhelashvili:2021eid, Dvali:2022fdv, Burgess:2023ifd}.
Thus, at low energies, both the PQ axion and the gauge axion describe the same physics.
However, the duality is broken  by the UV-completion.   

The PQ axion can be UV completed into the phase of a complex scalar field, with a spontaneously broken, anomalous symmetry, the $U(1)_{PQ}$ \cite{Peccei:1977hh,*Peccei:1977ur, Weinberg:1977ma, *Wilczek:1977pj}.
However, this UV-completion comes at the price of the quality problem.
  
On the other hand, the gauge axion, which has exact quality, has no known UV-completion within a renormalizable EFT. 
In fact, there are strong arguments indicating that the UV-completion must be linked to the UV-completion of gravity ~\cite{Dvali:2022fdv,  Dvali:2023llt, Dvali:2024dlb}.
While, from first glance this can be taken as a drawback, in reality it is a blessing, since the axion can be a direct window to the fundamental physics of gravity. 

In the present paper we shall obtain further indirect evidence for the fundamental gravitational origin of the gauge axion. 
Namely, by analysing the topological features of the vacuum we shall be lead to the conclusion that $B_{\mu\nu}$ is sourced by fundamental strings. 

\section{Domain Walls of the Gauge Axion}

Explicit solutions for Yang-Mills domain walls of the axion and the CS $3$-form have been discussed in the literature \cite{Dvali:1998ms, Dvali:1999pk,Dvali:2004tma}, \cite{Sakhelashvili:2021eid, Anber:2024gis}.
Here we are going to adjust this analysis to the present case. 

For this discussion, we consider the minimal model with pure QCD and the 2-form $B$, the gauge axion.
We ignore the quarks, or, since we are in the confining phase of QCD, the mesons.
From now on, we restore the scale $\Lambda$ and proper normalization factors.
The Lagrangian for this model is
\begin{align}
    \label{LagGA}
    \Lag = \Lambda^4\mathcal{K}\left(\frac{E}{\Lambda^2}\right) + \frac{m^2}{2}\tilde{C}^{\mu\nu\lambda}\tilde{C}_{\mu\nu\lambda},
\end{align}
where $\tilde C$ is given by \eqref{eq:Stuckelberg} and $m = \frac{\Lambda^2}{f}$ is the axion mass.
$\mathcal{K}(\frac{E}{\Lambda^2})$ is a generalized function of the 3-form field strength.  
As explained previously, we ignore the higher derivatives in $E$, and restrict to a bilinear term in $\tilde{C}$. 
An inclusion of higher order terms has no effect on the vacuum structure. 
It therefore only influences the core shape of topological defects but not their asymptotics.  
  
As shown in \cite{Dvali:2005an}, the above theory is dual to a pseudo-scalar axion $a$, with the potential determined by the relation  $V(a)' \propto \mathrm{inv}(\mathcal{K}')$.

As a specific example, we will consider the kinetic function discussed in \cite{Dvali:2005an},
\begin{align}
    \label{Ksine}
    \mathcal{K}\left(\frac{E}{\Lambda^2}\right) = \frac{E}{\Lambda^2}\arcsin{\left(\frac{E}{\Lambda^2}\right)}+\sqrt{1-\left(\frac{E}{\Lambda^2}\right)^2},
\end{align}
which corresponds to the axion potential 
\begin{align} 
    \label{sinegordon}
    V(a)=-\Lambda^4\cos\left(\frac{a}{f}\right) \,,
\end{align}
generated in the dilute instanton gas approximation. 
This potential leads to the well-known sine-gordon equation, which has a domain wall solution of the form (see, e.g., \cite{Vilenkin:1984ib}),
\begin{align}
    \label{profile}
    a_{DW}(x) = 4f\arctan\left(e^{mx}\right)\,,
\end{align}
where $x$ is a coordinate perpendicular to the wall.
Since eq. \eqref{Ksine} corresponds to the potential \eqref{sinegordon}, it is clear that there exists an equivalent domain wall solution in the language of the 3-form $C$.

In order to find this solution, let us consider the equation of motion that can be derived from eq. \eqref{LagGA}:
\begin{align}
    \Lambda^4 \partial_\mu\frac{\partial K}{\partial E}\epsilon^{\mu\nu\lambda\sigma} + m^2\tilde C^{\nu\lambda\sigma}=0.
\end{align}
We also keep in mind the generalized Proca condition \eqref{Proca1}.
We are interested in finding a static domain wall solution, which w.l.o.g. is chosen to lie in the $xy$-plane.
Therefore, the only non-zero component of $\tilde C$ (up to permutations) is $6 \tilde C^{012} \equiv f  J(z)$ and correspondingly $E = f J'(z)$. 
With this ansatz and eq. \eqref{Ksine}, the equation of motion can be reduced to
\begin{align}
   - \frac{J''}{\sqrt{1-\left(\frac{J'}{m}\right)^2}} +m^2J=0\,,
\end{align}
where we further used $m =\Lambda^2/f$.
This equation has the analytic solution 
\begin{align}
    \label{current}
    J = 4\frac{e^{mz}}{1+e^{2mz}},
\end{align}
which can be shown by direct computation.
We note that an analogous solution has been derived in \cite{Anber:2024gis}.

\begin{figure}
    \centering
    \includegraphics[width=0.8\linewidth]{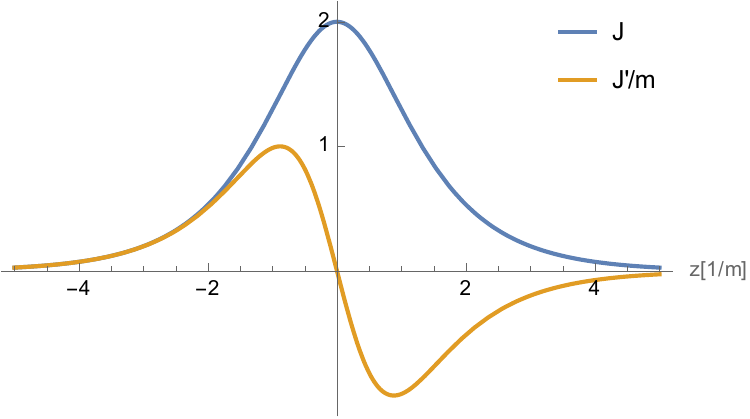}
    \caption{Plot of the domain Wall current $J(z)$ and its derivative $J'(z)$ (normalized w.r.t. $m$).}
    \label{fig:solution}
\end{figure}

This solution is equivalent to the domain wall solution of the sine-gordon potential \eqref{sinegordon}.
This can be seen from two arguments:

First, by construction, the kinetic function \eqref{Ksine} and the potential \eqref{sinegordon} are dual to each other. 
Second, $\tilde C$ is dual to $\partial_\mu a$ and it is straightforward to show that $a_{DW}' =\Lambda^2J(z)$. 
In the pseudoscalar description, we can define the topological current $J^{\mu\nu\lambda}_{DW}=f\epsilon^{\mu\nu\lambda\sigma}\partial_\sigma a$, which can be matched to the 3-form by the identity 
\begin{align}
    \label{matching}
    \tilde C^{\mu\nu\lambda} = \frac{1}{\Lambda^2}J_{DW}^{\mu\nu\lambda}.
\end{align}
This matching is also physically sensible, as both the 3-form and the current are topological quantities.
Therefore, the 3-form formalism describes the same axionic domain wall as the pseudoscalar picture, only in terms of the current instead of the axion profile.
Furthermore, notice that $E=fJ'(z)=\frac{1}{m}a_{DW}''(z)=-\Lambda^2\sin{(a_{DW}/f)}$, matching the duality relations needed to pass from one description to the other \cite{Dvali:2005an, Dvali:2017mpy}.

\section{Cosmic Strings of the Gauge Axion}

With the presence of domain walls in the gauge axion theory and their mapping on regular axionic domain walls established, it is now time to turn to the other type of defect that can appear in axion models: cosmic strings. 
In the PQ scenario these objects form due to the spontaneous breaking of the global $U(1)_{PQ}$-symmetry.

However, as previously mentioned, unlike the PQ axion, the gauge axion can not be UV-completed into the phase of a complex scalar. 
This means that within the EFT of $B_{\mu\nu}$ there exists no direct 
equivalent of a restored symmetry phase. 
Correspondingly, within the EFT there is no apparent phase transition with spontaneous symmetry breaking that could create axionic strings.

Nevertheless, we shall argue that the strings, which source the gauge axion must exist. 
The caveat is that they are not describable as solitons in low energy EFT and their resolution requires tools of the UV-theory. 
These strings, whose origin lies in the fundamental theory, will be referred to as fundamental strings.
We shall deduce their existence by analyzing axionic domain walls.

\subsection{The Solution as a Path in the Field Space}

Before doing so, we notice that the topological charge, given by the integral,

\begin{align}
    Q_C = \Lambda^2\int \epsilon_{\mu\nu\lambda\sigma}\tilde C^{\mu\nu\lambda} \d x^\sigma \,,
\end{align}
takes the value  $Q_C =2\pi f^2$ on the domain wall solution \eqref{current}.
This implies that the solution interpolates between two degenerate vacua. 

Now, in between these two vacua, $\tilde C^{\mu\nu\lambda}$ (and $E$) is well-defined everywhere on a solution. 
This means that we can describe the solution as a continuous non-singular path in the field space.

The second crucial ingredient is that the physics is periodic in $E$ (or equivalently in $G\tilde{G}$). 
Correspondingly, there must exist a completion of the above path, which leads back to the original vacuum. 

The only remaining question is, whether this closed path is covered by a single domain wall or a sequence of $N$ walls. 
In both cases, a string sourcing the gauge axion appears in one form or another. 

In other words, despite the fact that the cosmic strings of the gauge axion cannot be described as solitons of the low energy EFT, the duality between $B_{\mu\nu}$ and $a$, which is exact at the level of the low energy theory, allows to prove their existence, without the need to enter in the UV-completion.   

\subsection{Topological Configurations with Strings}

With the above knowledge, the full classification is mapped on the 
PQ string-wall systems (for a review, see e.g., \cite{Kim:1986ax}. 
For an updated discussion, including the effect of the QCD condensate on axionic string-walls, see \cite{Dvali:2025xur}). 
These are fully characterized by the anomaly-free discrete $Z_N$ symmetry. 

For $N=1$, the phase of the PQ field, $a/f$, interpolates between $0$ and $2\pi$ and the vacua on the two sides of the domain wall are identical, $a/f = a/f+2\pi$.

In contrast, for $N>1$, the phase interpolates between $\frac{2k\pi}{N}$ and $\frac{2(k+1)\pi}{N}$ ($N\in \mathbb N$, $0\leq k\leq N-1$), and the vacua across each elementary wall are different. 

Due to the low energy duality, this structure is fully inherited by the gauge axion walls.
For more clarity, let us consider the cases $N=1$ and $N>1$ separately. 

\subsubsection{$N=1$ Domain Walls}

Consider a planar infinite domain  wall in the case $N=1$.  
As mentioned above, the vacua on both sides of the wall are the same. In other words, the path describing a solution in the field space is closed. 
Correspondingly, it is possible to punch a hole in the domain wall and connect the two vacua.   

The edge of the hole is a cosmic string, around which the PQ phase winds. 
The field strength $E$ (and correspondingly $G\tilde{G})$) changes accordingly.  
This feature is there regardless of whether the axion is described by $a$ or $B_{\mu\nu}$. 
Correspondingly, the string exist in both cases. 
   
The appearance of the hole can also take place dynamically as a consequence of a tunneling process. 
From the point of view of the world-volume of the domain wall, the hole is a vacuum "bubble". 
Since the domain wall is a (2+1)-dimensional object, the boundary of the bubble is a (1+1)-dimensional object, looping to form a bubble. 
Therefore, the boundary of the domain wall is necessarily described by a cosmic string loop.  

\begin{figure}
    \centering
    \includegraphics[width=\linewidth]{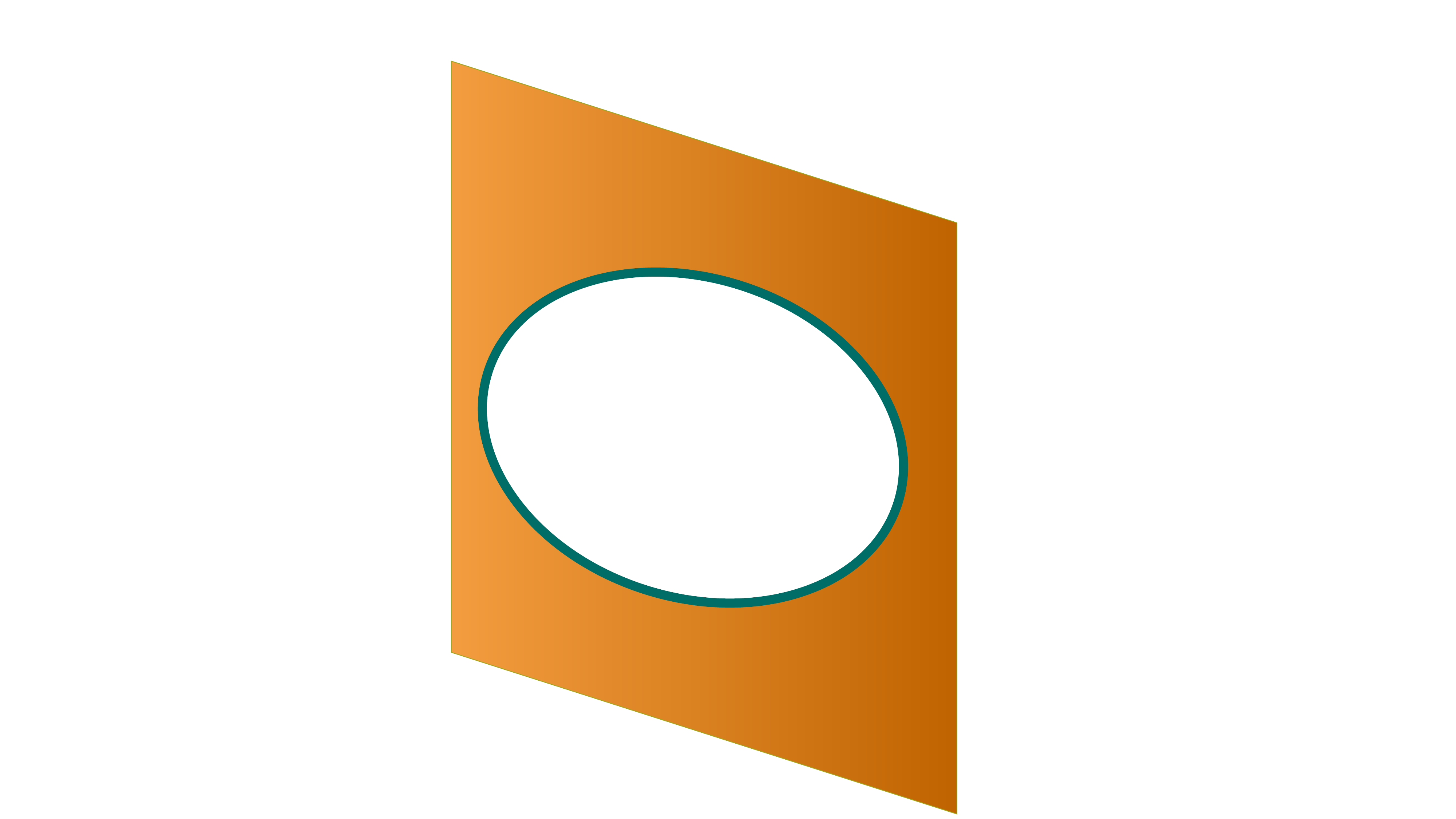}
    \caption{A domain wall with a hole in the middle. The cosmic string forms the boundary of the hole.}
    \label{fig:domainwallwithahole}
\end{figure}

\subsubsection{$N>1$ Domain Walls}

For $N>1$, the vacua across the wall are different. 
It is therefore not possible to punch a hole in the domain wall. 
Nevertheless, there exist wall configurations that make strings necessary. 
These are given by the junctions of $N$ walls.  
After the excursion across $N$ walls the path in the field space is closed. 
  
We can map this closed path to a closed path in a coordinate space. 
Since the neighboring vacua are separated by the domain walls, the path must necessarily enclose their junction.
This junction is a string. 

\begin{figure}
    \centering
    \includegraphics[width=\linewidth]{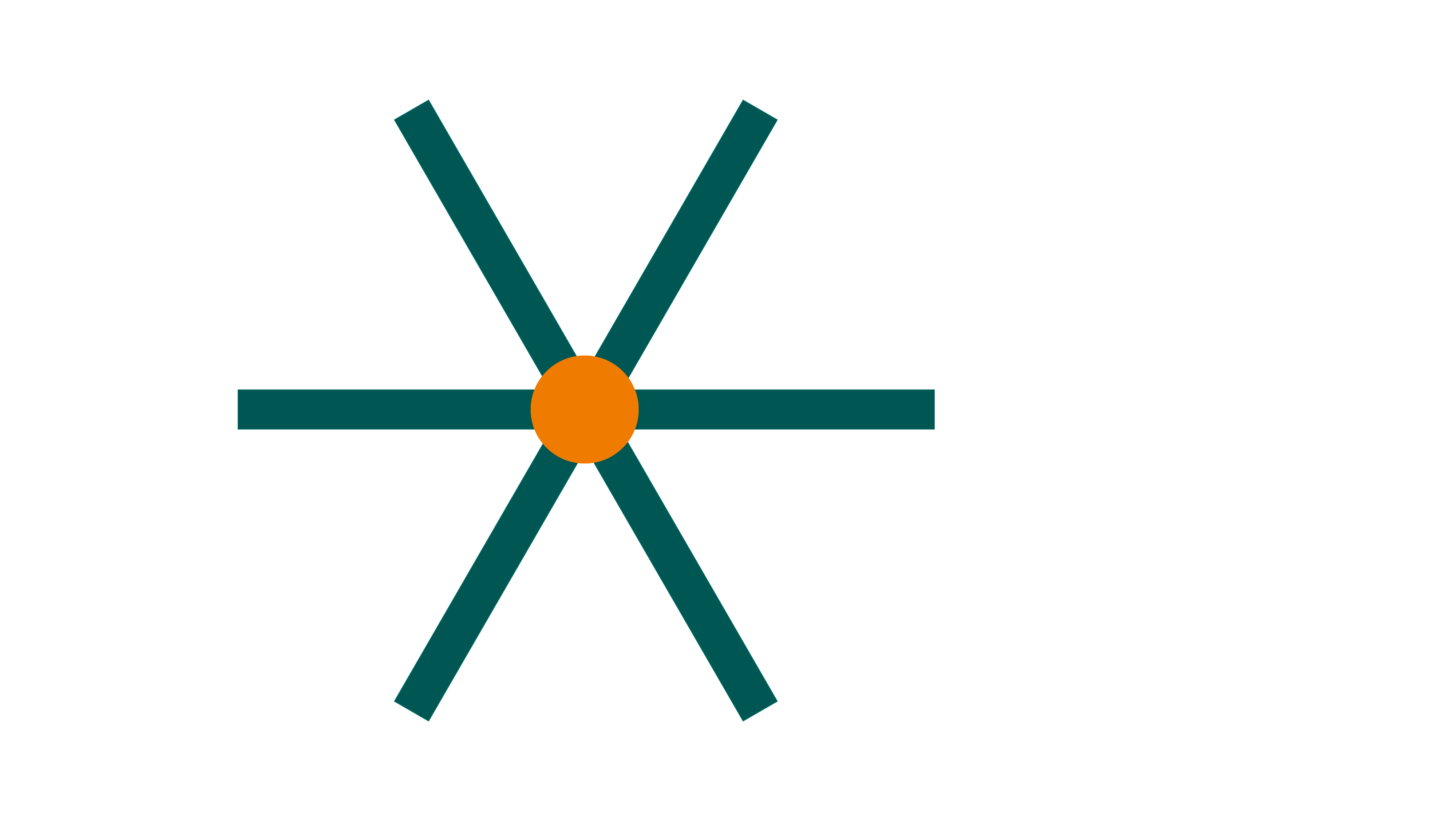}
    \caption{A string as the junction between several domain walls.}
    \label{fig:domainwalljunction}
\end{figure}

In the language of the PQ axion, the  
anomaly-free $Z_N$-subgroup of $U(1)_{PQ}$ can emerge from the axion coupling to $N$ quark 
flavors. 
For the case of the gauge axion, the origin of $N$ does not need to be specified. 
Rather, we treat it as a parameter to describe the structure of the string-wall system. 

\subsection{Properties of the Strings}

In the PQ mechanism, the cosmic strings are a  consequence of the spontaneous breaking of the $U(1)_{PQ}$-symmetry.
They are topological objects, described by classical solutions of the equations of motion of the PQ field. 

This is different for the strings of the gauge axion. 
Since the gauge axion cannot be UV completed into a complex scalar, we cannot 
describe the gauge axion strings as topological defects in the EFT. 
Instead, these strings are objects originating from the underlying fundamental theory. 
At the level of EFT, the strings can only be described by features which do not require the microscopic resolution of their structure. 

Interestingly, the necessity of domain walls, which are attached to the strings, already emerges at the level of the EFT as a consequence of the QCD gauge invariance. 
Indeed, in  order to account for the gauge symmetry, the shift of $B_{\mu\nu}$ \eqref{shift} in the string coupling, 
\begin{align}
    f\int_{1+1} \d X^{\mu} \wedge \d X^{\nu}\, B_{\mu\nu} \,, 
\end{align}
must be compensated by the shift \eqref{ShiftQCD} of $C_{\mu\nu\alpha}$ in the coupling to the wall, 
\begin{align}
    \int_{2+1} \d X^{\mu} \wedge \d X^{\nu}\, \wedge \d X^{\alpha} \, C_{\mu\nu\alpha } \,, 
\end{align}
showing that the $B$-string is necessarily a boundary of the $C$-wall \cite{Dvali:2006az}
\footnote{From the symmetry perspective one may consider the possibility that the string contains a world-sheet scalar degree of freedom, $\kappa$, which enters the coupling \cite{Dvali:2006az}, 
\begin{align*}
    \int d\sigma^2 (\epsilon^{ab} \partial_aX^{\mu}
    \partial_bX^{\nu}B_{\mu\nu} + \kappa) ,
\end{align*}
compensating the gauge shift of $B_{\mu\nu}$.
However, one has to keep track of the equation of motion for $\kappa$.
Even if such a possibility can be consistently realized, it is irrelevant for our analysis, since we are proving the existence of strings based on the existence of walls.}.
Notice that in the special case that a string possesses a conserved $U(1)$ Noether current $J_{\sigma}$, the following gauge invariant coupling can be written, 
\begin{align}
     \epsilon^{\mu\nu\lambda\sigma}\tilde C_{\mu\nu\lambda}J_{\sigma}\,.
\end{align}
This in principle allows to couple the gauge axion to Nielsen-Olesen type solitonic strings.
Of course, this coupling has to emerge as an effective coupling from the fundamental theory as well.  

Needless to say, the nature of the fundamental strings sourcing the gauge axion is expected to be very different from ordinary PQ strings. 
In particular, their creation may not be describable by an ordinary Kibble mechanism and instead requires scenarios involving a fundamental theory.  
For example, the creation of fundamental $F$ and $D$ strings in the context of brane inflation \cite{Dvali:1998pa} has been discussed in several papers \cite{Sarangi:2002yt, Dvali:2003zj}.  
More recent discussions on embedding the axion in brane constructions can be found in~\cite{Reece:2025thc, Niedermann:2025xnw}. 

Due to the above, the early cosmology of the gauge axion might be entirely different from the PQ cosmology. 
This will be described elsewhere \cite{gaugeaxion2}.

Finally, we would like to remark on the point recently made in \cite{Dvali:2025xur}, about the role of the QCD chiral quark condensate for an axionic string-wall system. 
In this paper it was argued that in many cases, the winding of the phase of the QCD quark condensate takes place around axionic strings, due to its mixing with the axion. 
Since the effect comes entirely from the long-distance regime, in which the duality between $a$ and $B_{\mu\nu}$ holds, we expect that the same effect must take place for the gauge axion strings, despite their fundamental origin.  

\section{Conclusion and Outlook}
In this paper, we have discussed the understanding of the strong-$CP$ problem from the TSV and its solution by the gauge axion \cite{Dvali:2005an}, which has exact quality. 
After  proving a direct connection between the TSV and the physicality of $\theta$-term, we have shown how the gauge axion makes it unphysical. 

Besides nullifying the problem to all orders in the operator expansion, the gauge axion serves as an explicit monitoring tool to track the $\theta$-vacua in various regimes, by the power of gauge redundancy. 
In particular, we have shown how the $\theta$ vacua re-emerge in the limit in which the gauge axion is decoupled.  
  
It has already been suggested \cite{Dvali:2022fdv} that, unlike the PQ axion, which suffers from the quality problem, the gauge axion must be UV completed directly in a fundamental theory of gravity. 
In the present paper  we provide new supporting evidence to this statement by observing that the strings which source the gauge axion must be fundamental.  
  
In arriving to the existence of strings, we relied on the fact that the gauge axion domain walls are fully describable as soliton solutions in the EFT, which can be dualized into analogous domain walls of a PQ type axion. 

Furthermore, we have argued that the presence and nature of these domain wall solutions makes the presence of strings unavoidable. 
Depending on whether the vacua across a domain wall are the same or not, they appear either as boundaries of holes, which can be punched into the walls, or as junctions connecting multiple domain walls.

However, unlike the case of the PQ strings, the strings sourcing the gauge axion are not describable as solitons of the low energy EFT and in this sense are fundamental. 
This implies that the gauge axion, which solves the strong-$CP$ problem, at the same time, represents a probe of fundamental strings. 

Of course, at the level of our bottom-up discussion, we cannot prove that these "fundamental" strings are necessarily the ones of a known superstring theory. 
However, the general indications are rather striking. 
    
This observed feature of the gauge axion, resonates with previous suggestions \cite{Dvali:2022fdv,  Dvali:2023llt, Dvali:2024dlb} that an axion of exact quality is required by the consistency of the UV-theory of gravity. 

Finally, our findings indicate that the early cosmology of the gauge axion is expected to be quite different from the PQ axion. 

{\bf Acknowledgments} \\
This work was supported in part by the Humboldt Foundation under Humboldt Professorship Award, by the European Research Council Gravities Horizon Grant AO number: 850 173-6,
by the Deutsche Forschungsgemeinschaft (DFG, German Research Foundation) under Germany's Excellence Strategy - EXC-2111 - 390814868, and Germany's Excellence Strategy under Excellence Cluster Origins. \\
 
 Disclaimer: Funded by the European Union. Views
and opinions expressed are however those of the authors
only and do not necessarily reflect those of the European Union or European Research Council. Neither the
European Union nor the granting authority can be held
responsible for them.

\appendix
  
\bibliographystyle{utphys}
\bibliography{etawalls.bib}		

\end{document}